# REMOTE SENSING OF VEGETATION DYNAMICS IN AGRO-ECOSYSTEMS USING SMAP VEGETATION OPTICAL DEPTH AND OPTICAL VEGETATION INDICES

*M. Piles[1], D. Chaparro[2], D. Entekhabi[3], A. G. Konings[4], T. Jagdhuber[5], G. Camps-Valls[1]*

[1]Image Processing Lab (IPL), Universitat de València, 46980 València, Spain.
[2]Universitat Politècnica de Catalunya, IEEC/UPC, Jordi Girona 1-3, 08034 Barcelona, Spain.
[3]Department of Civil and Environmental Engineering, Massachusetts Institute of Technology, Cambridge, MA 02139, USA
[4]Department of Earth System Science, Stanford University, Stanford, CA 94305, USA
[5]German Aerospace Center, Microwaves and Radar Institute, P.O. Box 1116, 82234 Wessling

## ABSTRACT

The ESA's SMOS and the NASA's SMAP missions, launched in 2009 and 2015, respectively, are the first two missions having on-board L-band microwave sensors, which are very sensitive to the water content in soils and vegetation. Focusing on the vegetation signal at L-band, we have implemented an inversion approach for SMAP that allows deriving vegetation optical depth (VOD, a microwave parameter related to biomass and plant water content) alongside soil moisture, without reliance on ancillary optical information on vegetation. This work aims at using this new observational data to monitor the phenology of crops in major global agro-ecosystems and enhance present agricultural monitoring and prediction capabilities. Core agricultural regions have been selected worldwide covering major crops (corn, soybean, wheat, rice). The complementarity and synergies between the microwave vegetation signal, sensitive to biomass water-uptake dynamics, and optical indices, sensitive to canopy greenness, are explored. Results reveal the value of L-band VOD as an independent ecological indicator for global terrestrial biosphere studies.

## 1. INTRODUCTION

The first two space missions operating at microwave L-band or "the water frequency channel" have been launched in the last decade to globally measure the Earth's surface soil moisture: the ESA's Soil Moisture and Ocean Salinity (SMOS, 2009-2017) and the NASA's Soil Moisture Active Passive (SMAP, 2015-2018). Unlike traditional optically-based sensing technologies, microwaves are not affected by atmospheric conditions and total coverage of the Earth's surface is obtained every 2-3 days.

Recent efforts have been dedicated to the development of an inversion strategy for SMAP – the multi-temporal dual-channel retrieval algorithm (MT-DCA)- that allows deriving information of vegetation water content and structure alongside soil moisture from dense and sparse vegetation-covered surfaces, including croplands [1,2]. Importantly, the method does not rely on ancillary optical information that could influence the spatial and temporal patterns of the microwave vegetation retrievals [3]. This new sensing technique is used in this study to derive microwave vegetation optical depth (VOD) from the first full year of SMAP data (April 1, 2015 to March 31, 2016). Although VOD is known to be sensitive to above ground biomass and plant water content, regional studies focused on specific biomes (e.g. dry and wet tropical forests, agricultural lands, wetlands) are required to fully understand the useful information on the canopy status that can be derived from this observational data set.

This study investigates how SMAP VOD matches measures and statistics of crop phenology in a selection of major core agricultural regions. The complementary information provided by new microwave and traditionally-used optical sensors in agro-ecosystems is explored as the basis for a potential multi-sensor technique for global cropland mapping.

This research was supported by the Spanish Ministry of Science and Innovations through the National R+D Plan by means of the project PROMISES ESP2015-67549-C3 by the NASA Soil Moisture Active Passive project and the EU under the ERC consolidator grant SEDAL-647423.

## 2. PROPOSED APPROACH

The temporal dynamics of SMAP VOD (2-3 days, 36-km) are compared to those of the MODIS Enhanced Vegetation Index (EVI), contained in the MOD13C1 product (16-day composite, 0.05º). The EVI is a version of the commonly used Normalized Difference Vegetation Index (NDVI), also related to chlorophyll presence and plant greenness but improved to reduce atmospheric effects and avoid the saturation evident in NDVI products [4]. SMAP soil moisture and GPM precipitation measurements (1-day, 0.01º) are also used as supporting information.

MODIS EVI and GPM precipitation data are first aggregated to the SMAP EASE2 36-km grid by area-overlap weighted averaging. Then, regional time-series of VOD, EVI, soil moisture and precipitation are obtained for the selected target regions (see Table 1). All target regions enclose 36 km pixels with (at least) 90 % croplands, according to the 3 km MODIS IGBP land cover classification.

The correspondence of VOD time series with crop progress survey data reported by the United States Department of Agriculture (USDA) over Iowa is analyzed. Correlations of VOD with percentage of crop progress during growth-flowering & maturity stages are calculated.

Table 1. Target regions: location, crop type(s) and number of 36 km pixels enclosed

|  | Crops | N |
|---|---|---|
| Iowa | Corn and Soybean | 72 |
| Argentina | Soybean | 45 |
| SW Australia | Wheat | 18 |
| Nigeria | Mixed crops | 36 |
| Senegal | Mixed crops | 18 |
| East Thailand | Rice | 35 |
| West Thailand | Rice | 18 |
| North India | Rice and wheat | 45 |

## 3. RESULTS AND DISCUSSION

The phenology of corn and soybean fields in Iowa is shown in Fig. 1a. In the top plot, it can be seen that VOD and EVI are in phase, and that the peak of VOD is delayed with respect to that of EVI. This can be explained by the fact that, after the crop reaches maximum greenness, there is still a process of growing/changing structure that is captured by the microwave signal. The rise of VOD after harvest is probably due to ground disturbance (i.e. soil tilling) or imperfect detection of ice/snow.

The phenology of soybeans in Argentina is shown in Fig. 1b. Results are consistent with those obtained for Iowa: VOD and EVI are in phase, with the peak of biomass coming after the peak of photosynthesis.

Over Nigeria (Fig 1.c), it can be seen that EVI is in phase with precipitation and soil moisture, and the VOD peak comes after the end of the raining season.

As expected, the effect of standing water on rice fields over West Thailand (Fig. 1d) is evident for the VOD signal: it remains low while rice fields are inundated and peaks after they are drained. In this case, EVI is in phase with soil moisture but is out-of-phase with VOD.

Results over SW Australia, Senegal, East Thailand and North India confirm that crop biomass and greenness are in phase except when there is standing water (rice fields), and that VOD is delayed with EVI. Also, the VOD peak magnitude reflects crop biomass, with approximate values of 0.25 for soybean, 0.3 for wheat, and 0.3-0.4 for corn and rice.

A direct comparison of VOD dynamics with crop phenological stages provided by the United States Department of Agriculture (USDA) has been performed for the four districts within the Iowa target region. Results show a good agreement of VOD with crop progress data: the timing of the VOD peak coincides with the end of the flowering (corn silked, soy blooming) and the start of the maturity stages (corn dented, soy turning), and the timing of minimum VOD coincides with the harvesting dates. Sample results are shown in Fig. 2 for the central west district. Results over the north west, north and central districts are consistent (not shown). A good relationship was also found between smoothed SMOS VOD data and crop stages in Iowa in a recent study [5]. These results confirm the potential of the two L-band missions for monitoring the growth and development of crops.

Further analysis of key stages show that the reproductive stages (corn silked, soy blooming) in mixed corn-soy fields in Iowa are highly correlated with increasing VOD ($R^2 = 0.6$-$0.7$), reflecting the gain on plant water content and biomass in these crop phases. Likewise, the maturation of corn and soy (corn dented and mature, soy turning and shedding), consisting of drying of plants and fruits, is correlated with the decreasing VOD signal ($R^2 = 0.5$-$0.7$).

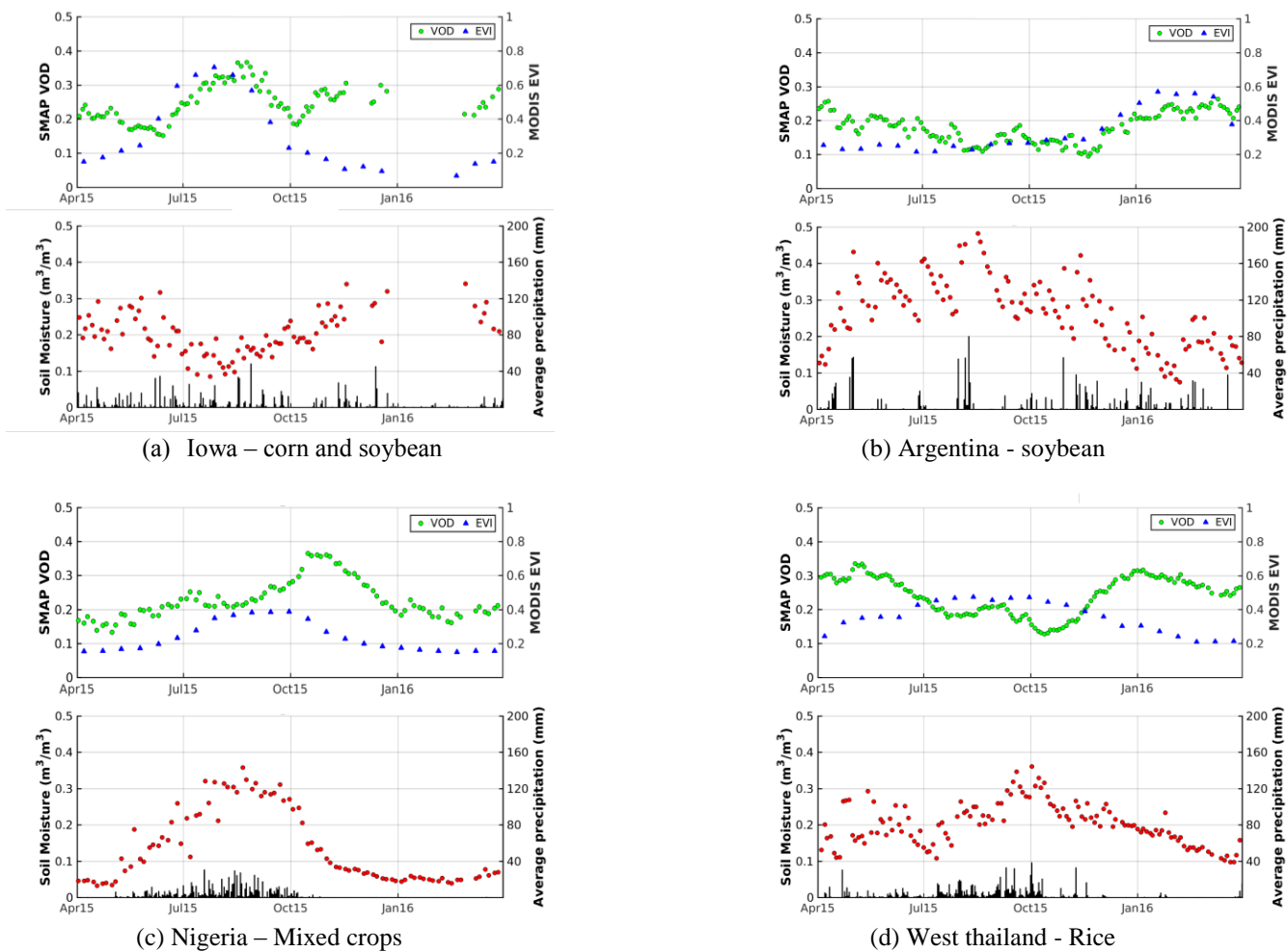

**Figure 1.** Time series of SMAP VOD (green dots), MODIS EVI (blue triangles), SMAP soil moisture (red dots) and GPM precipitation (black vertical bars) for (a) Iowa, (b) Argentina, (c) Nigeria and (d) West Thailand target regions.

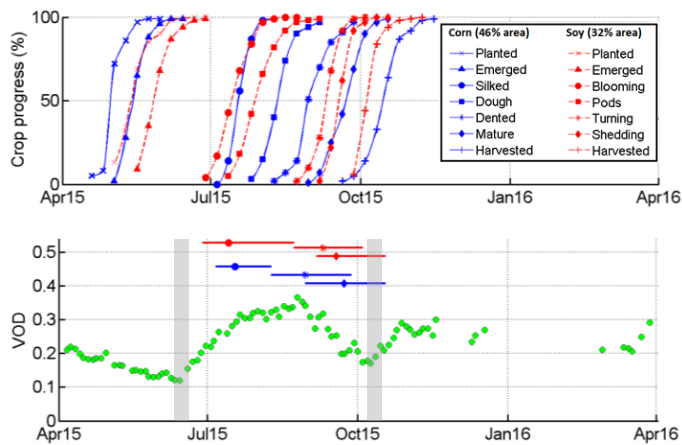

**Figure 2.** Top: crop progress information for corn (blue) and soybean (red) in the west-central district of Iowa for the study period. Bottom: SMAP VOD time series. Grey vertical bars indicate the start (emerging) and end (harvesting) of the crop phenologycal cycle. Horizontal bars mark the progress of of flowering (corn silked, soy blooming) and maturity (corn dented and mature, soy turning and shedding) stages

Although vegetation indices are not intrinsic physical quantities, they are widely used to indirectly observe vegetation dynamics. EVI is an optically-based measure of vegetation canopy greenness, a composite property of leaf chlorophyll, leaf area, canopy cover, and canopy architecture. Our results support that L-band microwave VOD measurements can complement the optically based greenness measurements by providing direct and immediate information about plant water content and above ground biomass. Results indicating the potential of a multi-sensor technique for global cropland mapping will be presented at the time of the conference.